\documentstyle[]{article}

\def\mypagenumber{1}
\def\mydate{4 September 1998}
\def\myend{\end{document}}

\date{}
\newcommand{\beeq}{\begin{equation}}
\newcommand{\eneq}{\end{equation}}
\newcommand{\beqn}{\begin{eqnarray}}
\newcommand{\eeqn}{\end{eqnarray}}

\def\mybig{\displaystyle \strut }

\def\dd{\partial}
\def\la{\raise.16ex\hbox{$\langle$} \, }
\def\ra{\, \raise.16ex\hbox{$\rangle$} }
\def\ran{\raise.16ex\hbox{$\rangle$} }
\def\go{\rightarrow}

\def\onehalf{ \hbox{${1\over 2}$} }

\def\psibar{ \psi \kern-.65em\raise.6em\hbox{$-$}\lower.6em\hbox{} }
\def\tpsi{ \widetilde\psi}
\def\trho{\tilde\rho}
\def\tM{\widetilde M}
\def\tpsibar{ \tpsi \kern-.65em\raise.9em\hbox{$-$}\lower.9em\hbox{} }
\def\zetabar{ \zeta\kern-.65em\raise.6em\hbox{$-$}\lower.6em\hbox{} }
\def\mbar{ m \kern-.78em\raise.4em\hbox{$-$}\lower.4em\hbox{} }

\def\Bbar{ B \kern-.73em\raise.6em\hbox{$-$}\hbox{} }

\def\L{ {\cal L} }
\def\ep{\epsilon}

\def\vphi{ {\varphi} }

\def\tchi{{\tilde \chi}}
\def\tphi{\tilde \phi}
\def\tPhi{\tilde\Phi}
\def\tq{\tilde q}
\def\tQ{\tilde Q}

\def\myfrac#1#2{{\mybig #1\over \mybig #2}}

\def\boxit#1{$\vcenter{\hrule\hbox{\vrule\kern3pt
     \vbox{\kern3pt\hbox{#1}\kern3pt}\kern3pt\vrule}\hrule}$}
\def\bigbox#1{$\vcenter{\hrule\hbox{\vrule\kern5pt
     \vbox{\kern5pt\hbox{#1}\kern5pt}\kern5pt\vrule}\hrule}$}

\def\@makefnmark{{$\!^{\@thefnmark}$}}

\begin{document}

\setcounter{page}{\mypagenumber}

{\baselineskip=10pt \parindent=0pt \small
{\mydate ~(corrected) \hfill Appeared in J. Phys. A\\} 
\rightline{cond-mat/9707129}
}

\vspace{10mm}

\centerline {\Large\bf Gauge Theory Description of Spin Ladders}

\vspace*{5mm}
\centerline {\normalsize  
Yutaka Hosotani\footnote{e-mail: yutaka@mnhepw.hep.umn.edu}}

\centerline {\small\it School of Physics and Astronomy, University
       of  Minnesota}
\centerline {\small\it Minneapolis, Minnesota 55455, U.S.A.}

\vspace*{3mm}
\normalsize

\begin{center}
{\begin{minipage}{4.8truein}
                 \footnotesize
                 \parindent=0pt 
A $s$=$\onehalf$ antiferromagnetic spin chain is equivalent to the two-flavor
massless Schwinger model in an uniform background charge density in the
strong  coupling.  The gapless mode of the spin chain is represented by a 
massless boson of the Schwinger model.  In a two-leg spin ladder system
the massless boson aquires a finite  mass due to inter-chain interactions.
The gap energy is found to be about  $ .25 k \, |J'|$
when the inter-chain Heisenberg coupling $J'$ is  small compared with
the intra-chain Heisenberg coupling.  $k$ is a constant of O(1).  It is
also shown that  a  cyclically symmetric $N_\ell$-leg ladder system is
gapless or gapful for an odd or even $N_\ell$, respectively.
\par
                 \end{minipage}}\end{center}

\vskip .5cm



A $s=\onehalf$ spin chain with antiferromagnetic nearest-neighbour
Heisenberg couplings is exactly solved by the Bethe ansatz.\cite{Bethe}
It has a gapless excitation.  A two-leg spin ladder consists of
two spin chains  coupled each other.  Experimentally  a two-leg spin
ladder  system has no gapless excitation.\cite{exp1,exp2,exp3,exp4}  The
gapless mode of spin chains becomes gapful.  In this paper we give, 
without resorting to numerical evaluation, a deductive microscopic argument
which shows why and how it happens.

Spin ladder systems are not exactly solvable.  Various approximation
methods have been employed in the
literature.\cite{Rice,theory,numerical,Sierra,Schulz,bosonize}  We first
show that a
$s=\onehalf$ spin chain is equivalent to the two-flavor massless Schwinger
model in the strong coupling in an uniform background charge density.  The
two-flavor Schwinger model has a massless boson excitation, which
corresponds to the gapless excitation in the Bethe ansatz.
A spin ladder system
is described as two sets of two-flavor Schwinger models which interact
with each other by four-fermi interactions.  

An antiferromagnetic spin chain is described by
\beeq
H_{\rm chain}(\vec S) = J \sum \vec S_n \cdot \vec S_{n+1}  
\qquad (J>0)
\label{spinchain1}
\eneq
wheras a two-leg spin ladder is described by
\beqn
H_{\rm ladder}(\vec S, \vec T) 
  &=& H_{\rm chain}(\vec S) + H_{\rm chain}(\vec T) 
     + H_{\rm rung}(\vec S, \vec T)  \cr 
\noalign{\kern 5pt}
H_{\rm rung}(\vec S, \vec T) &=&  J' \sum  \vec S_n \cdot \vec T_n ~.
\label{spinladder1}
\eeqn
Consider first a $s=\onehalf$
anti-ferromagnetic spin chain $H_{\rm chain}(\vec S)$.
  We express the spin operator in terms of electron
operators by $\vec S_n = c_n^\dagger \onehalf \vec \sigma c_n^{}$.
With the aid of the Fierz transformation we have
$4 \vec S_n \cdot \vec S_{n+1}= - \{ c_n^\dagger c_{n+1}^{} ,
c_{n+1}^\dagger c_n^{} \} + 1 
- (c_n^\dagger c_n^{} -1)(c_{n+1}^\dagger c_{n+1}^{} -1)$.
The last term drops as the half-filling condition $c_n^\dagger c_n^{} =1$
is satisfied for a spin chain.  The first term is linearized by
introducing an auxiriary field, or by the Hubbard-Stratonovich
transformation.  The Hamiltonian $H_{\rm chain}$ is equivalent to the
Lagrangian
\beeq
L^1_{\rm chain} = \sum \Big\{ ~ i \hbar c_n^\dagger \dot c_n^{} 
   - \lambda_n (c_n^\dagger c_n^{} - 1)    
-{J\over 2} (U_n^* U_n - U_n^{} c_n^\dagger c_{n+1}^{} - 
U_n^* c_{n+1}^\dagger c_n^{} ) 
~ \Big\} ~.
\label{spinLagrangian1}
\eneq
$U_n$ is a link variable, defined on the link connecting
sites $n$ and $n+1$.  $\lambda_n$ is a Lagrange multiplier enforcing
the half-filling condition at each site.  The transformation is valid
for $J>0$.   The Lagrangian $L^1_{\rm spin}$ has local $U(1)$ gauge
invariance.

We consider a periodic chain of $N$ sites: $\vec S_{N+1}= \vec S_1$.
The mean field energy is evaluated, supposing $|U_n| = U$,
to be $E_{\rm mean} = J \{ \onehalf N U^2 -  2U \cot(\pi/N) + 
{1\over 4} N \}$.  For large $N$ it has a sharp minimum at $U=2/\pi$.
Radial fluctuations of $U_n$'s are suppressed, though quantum
fluctuations of the phase of $U_n$'s cannot be neglected.
We write
\beeq
U_n = {2\over \pi} \, e^{i \ell A_n}
\label{link}
\eneq
where $\ell$ is the lattice spacing.  We need to  incorporate quantum
fluctuations of $\lambda_n$ and $A_n$ to all orders.
With (\ref{link}) substituted
the Lagrangian (\ref{spinLagrangian1}) becomes that of lattice
electrodynamics.  

To make this point clearer, we take the continuum limit.  For an
anti-ferromagnetic spin chain, two sites form one block. The even-odd
site index becomes an internal (spin) degree of the Dirac field in the
continuum limit.  The correspondence is given by
\beeq
\cases{\psi_1^{(a)}(x) 
  = \myfrac{(-i)^{2s-1}}{\sqrt{2\ell}} ~ c_{2s-1,a} &at odd site\cr
\psi_2^{(a)}(x)
  = ~\myfrac{(-i)^{2s}}{\sqrt{2\ell}} ~~ c_{2s,a}&at even site\cr}
\label{correspondence1}
\eneq
where $x$ corresponds to $(2s-1)\ell$ and $2s\ell$.  With the given 
normalization   
$\{ \psi_j^{(a)}(x) , \psi_k^{(b)}(y)^\dagger \} 
= \delta^{ab} \delta_{jk} \delta_L(x-y)$ in the continuum limit,
where $\delta_L(x)$
is the periodic delta function with the period $L=N\ell$.  The phase 
factors in (\ref{correspondence1}) reflect  the fermi momentum 
$k_F =\pm \onehalf \pi$ at the half filling.

The term $\sum_n c_n^\dagger c_{n+1}^{} + {\rm h.c.}$ becomes
$ 2i\ell \sum_a \int dx \, 
\big( \psi_1^{(a)\dagger}\dd_x\psi_2^{(a)}
+ \psi_2^{(a)\dagger}\dd_x\psi_1^{(a)} \big) $.
Hence in the continuum limit the original spin Hamiltonian
(\ref{spinchain1}) is transformed to a system with the Lagrangian density
\beeq
\L^2_{\rm chain}[A_\mu, \psi] = - {1\over 4e^2} \, F_{\mu\nu}^2
+ \sum_{a=1}^2 c \psibar^{(a)}  \gamma^\mu 
  \Big( i \hbar \dd_\mu - {1\over c} A_\mu \Big)
\psi^{(a)} + {1\over \ell} A_0  ~.
\label{Schwinger1}
\eneq
Here the Dirac matrices are  $\gamma^0=\sigma_3,
\gamma^1=i\sigma_2$. The ``light'' velocity $c$ is given by
$c = 2 \ell J/\pi\hbar$.   $x_0 = c t$ and $(A_0, A_1) = (\lambda, c A)$.
Although the Maxwell term is absent in the $\ell \go 0$ limit,
it is generated at finite $\ell$.  The coupling constant
$e$ must be expressed in terms of $J$ and $\ell$.  From the dimensional
analysis $e^2 = k^2 J/\ell$ where $k$ is a constant of O(1).  Note that
in the $\ell \go 0$ limit with $c$ kept fixed, $e^2$ diverges as
$\ell^{-2}$.

This is nothing but the two-flavor massless Schwinger model in the 
strong coupling in an uniform background charge density.  The  term
$A_0/\ell$ representing the background charge arises from the half-filling
condition.  The system is neutral as a whole.

Notice that the spin index $a$ of original electrons becomes
a flavor index in (\ref{Schwinger1}), while the even-odd index $j$
becomes a spin index of the Dirac field $\psi^{(a)}(x)$.
The two-flavor nature reflects the electron  spin $\onehalf$.

The correspondence of the spin chain model to QED$_2$ has been
noted in the literature, but the rigorous derivation has not
been given  before.\cite{Itoi} In particular, the importance of the two-flavor
nature has not been recognized.  Mapping to $SU(2)$ gauge theory also
has been suggested.\cite{Murdy}

The two-flavor massless Schwinger model is exactly
solvable.\cite{Schwinger}  Quantum  fluctuations of all fields,
$\psi^{(a)}$ and $A_\mu$, can be completely taken into account.  With
the periodic boundary condition, the model is two-flavor QED$_2$ defined
on a circle, which has been  analysed in detail by the bosonization
method.\cite{Nakawaki,Hosotani1,Hosotani2}

The bosonization formula for the left and right moving components
of the Dirac fields is
\beqn
\psi^a_\pm(t,x) = {\mybig 1\over\mybig \sqrt{L}} 
\, C^a_\pm \,
 e^{\pm i \{ q^a_\pm + 2\pi p^a_\pm (t \pm x)/L \} }
  N_0[ e^{\pm i\sqrt{4\pi}\phi^a_\pm (t,x) } ] \quad (a=1,2)
\label{bosonization}
\eeqn
where  $C^a_+=e^{i\pi \sum_{b=1}^{a-1}(p^b_++p^b_-)}$ 
and $C^a_-=e^{i\pi\sum_{b=1}^a(p^b_+-p^b_-)}$.   
$\phi^a_+$ ($\phi^a_-$) represents left (right) moving modes.  $N_0[~]$
denotes the normal ordering in a basis of massless fields.
The Hamiltonian  becomes\cite{Hosotani2}
\beqn
&&H^2_{\rm chain} = {e^2 L\over 2} P_W^2 + 
\sum_{a=1}^2 {\pi\hbar c\over 2L} \Big\{ Q_a^2 + \big( Q_{5a} + 
{\Theta_W\over \pi} \big)^2 \Big\} \cr
\noalign{\kern 6pt}
&&\hskip 1.5cm + \int_0^L dx \, {\hbar c\over 2} 
  \big( {1\over c^2} \dot\Phi^2 + \Phi'^2 
               + {2e^2\over \pi\hbar c}  \Phi^2  
+ {1\over c^2} \dot\chi^2 + \chi'^2 \big)   ~.
\label{bosonizeH1}
\eeqn
The neutrality condition reads $Q_1 + Q_2 = L/\ell = N$.
$\Theta_W$ and $P_W$ are the Wilson line phase $e^{i\Theta_W}= 
\exp \big[ (i/\hbar c) \int_0^L dx \, A_1 \big]$ and its conjugate
momentum.
$Q_a = -p_a^+ + p_a^-$ and $Q_{5a}=p_a^+ + p_a^-$ are charge and axial
charge of the
$a$-th flavor, respectively, both of which take integer eigenvalues and
commute with the Hamiltonian.  $\Phi = (\phi_1 + \phi_2)/\sqrt{2}$ and
$\chi = (\phi_1 - \phi_2)/\sqrt{2}$ where $\phi_a=\phi^a_++\phi^a_-$
and $\int dx \, \phi_a = 0$.  

The $\Phi$ field has  a Schwinger mass $\mu$ where $\mu^2=2 e^2\hbar/\pi
c^3$.   The excitation energy is $\mu c^2 = 2 k  J /\pi\sim .637
kJ$.
 The $\chi$ field is massless, which 
corresponds to the gapless excitation in the spin chain and controls
the behavior of correlation functions at large distances.
The  wave function for the zero mode part is written  as
\beqn
&&|\Psi\ra = \sum_{n,r} \int dp_W  \, |p_W, n, r \ra
   \, e^{-ir\vphi + 2\pi inp_W} f(p_W, \vphi+\pi p_W) \cr
&&P_W |p_W, n, r\ra = p_W |p_W, n, r\ra \cr
\noalign{\kern 5pt}
&&p^a_\pm |p_W, n, r\ra 
  = (n +  r \delta_{a,1} \mp \hbox{${1\over 4}$} N ) |p_W, n, r\ra
\label{wavefunction1} 
\eeqn
where $f(p_W, \vphi)$ must solve the Schr\"odinger equation
\beqn
&&K(p_W,\vphi) \, f(p_W, \vphi)  = \ep \, f(p_W, \vphi)  \cr
\noalign{\kern 10pt}
&&K(p_W,\vphi) =  - {1\over \pi^2} {\dd^2\over \dd p_W^2} 
  - {\dd^2\over \dd \vphi^2}
 - \bigg( {\mu c L p_W\over 2 \hbar} \bigg)^2 ~.
\label{wavefunction2}
\eeqn
For the ground state $f(p_W, \vphi) ={\rm const} \cdot e^{-\pi\mu c L
p_W^2/4\hbar} $.

In the Schwinger model there is a $\theta$ parameter
characterizing states.  The wave function (\ref{wavefunction1})
corresponds to
$\theta=0$. The $\theta$ vacuum originates from the invariance under
large gauge  transformatins and the chiral anomaly in the continuum
theory.\cite{Hosotani1} In the lattice spin systems the lowest energy
state with
$\theta=0$ is expected to be singled out.

Employing the bosonization formula, the critical exponent of 
the spin-spin correlation function $\la \vec S(2n) \vec S(0) \ra
\sim n^{-\eta}$ ($n \gg 1$, $n \ll N$) is found to be $\eta=1$, 
which agrees with the result from the Bethe ansatz.

Now we consider a spin ladder system (\ref{spinladder1}).
In the absence of the inter-chain rung interaction ($J'$=0) the
system is equivalent to the two sets of two-flavor massless Schwinger
models  described by $\L_{\rm chain}^2 [A_\mu, \psi] 
+ \L_{\rm chain}^2 [\tilde A_\mu, \tpsi]$. With the aid of  the
correspondence (\ref{correspondence1}), the inter-chain interaction 
$H_{\rm rung}$ in the continuum limit is written as
\beeq
H_{\rm rung}^3 = {J' N\over 2}  + {\ell J'\over 4} \int dx \, \Big(
\{ \psi^{\dagger} \tpsi , \tpsi^\dagger \psi \} +
\psi^\dagger \psi \cdot \tpsi^\dagger\tpsi + 
\{ \psibar \tpsi , \tpsibar \psi \} + 
\psibar \psi \cdot \tpsibar\tpsi  \Big) 
\label{spinladder2}
\eneq
where every quantity in the expression is flavor singlet;
$\psi^\dagger \tpsi = \sum_{a=1}^2 \psi^{(a)\dagger} \tpsi^{(a)}$
etc.  Notice that both charge density and scalar density operators
appear in (\ref{spinladder2}).  The chiral symmetry is broken, which
leads to mass generation.

When expressed in terms of 
$\psi_\pm$ and $\tpsi_\pm$, $H_{\rm rung}^3$ contains many terms.  
The Hamiltonian is simplified in the large volume limit $L=N \ell\go \infty$.
Define $\rho_a=\psi^{(a)\dagger}\psi^{(a)}$, 
$M_a=\psi_+^{(a)\dagger}\psi_-^{(a)}$, and corresponding $\trho_a$
and $\tM_a$.  Relevant terms in $H_{\rm rung}^3$ are
\beqn
H_{\rm rung}^3  &\sim& H_{3a}+H_{3b} \cr
\noalign{\kern 10pt}
H_{3a} &=& {J'\ell\over 4} \int dx \, (\rho_1 -
\rho_2)(\trho_1-\trho_2) \cr
\noalign{\kern 10pt}
H_{3b} &=&  {J'\ell\over 4} \int dx \,
\Big\{  (M_1-M_2)(\tM_1^\dagger-\tM_2^\dagger) + ({\rm h.c.}) \Big\} ~.
\label{spinladder3}
\eeqn
Terms of the form $M_a \tM_b$ are suppressed as fluctuations in $Q_a$ 
are small compared with the average $N/2$.

Boson fields associated with $\psi$ and $\tpsi$ are denoted by
$(\Phi,\chi)$ and $(\tPhi,\tchi)$, respectively. We introduce a new
orthonormal basis:
$\Phi_\pm=(\Phi \pm \tPhi)/\sqrt{2}$ and $\chi_\pm=(\chi\pm\tchi)/\sqrt{2}$.
The first term in (\ref{spinladder3}) is
\beeq
H_{3a} = {J'\ell\over 4L} (Q_1-Q_2)(\tilde Q_1-\tilde Q_2)
+ \int dx {J'\ell\over 4 \pi} 
         \Big\{ (\dd_x\chi_+)^2 - (\dd_x\chi_-)^2  \Big\} ~.
\label{spinladder4}
\eneq
It changes the propagation velocities of $\chi_\pm$ fields.

It follows from (\ref{bosonization}) that
\beeq
M_a^{} \tM_b^\dagger = e^{-2\pi i(Q_a - \tQ_b)x/L} e^{-i(q_a-\tq_b)}
{1\over L^2} N_0[e^{-i\sqrt{4\pi}(\phi_a-\tphi_b)}] ~.
\label{MMrelation1}
\eneq
Note that 
$N_0[e^{i\beta\chi}] = B(mcL/\hbar)^{\beta^2/4\pi} N_m[e^{i\beta\chi}]$ 
where the reference mass in the normal
ordering $N[~]$ is shifted from 0 to $m$. $B(0)$=1 and $B(z)\sim
e^\gamma z /4\pi$ for $z\gg 1$.\cite{Hosotani1}  That is, if all fields
become massive, (\ref{MMrelation1}) is nonvanishing in the $L\go\infty$
limit. Otherwise (\ref{MMrelation1}) vanishes.
In passing, terms not included in (\ref{spinladder3}) are
suppressed exponentially in the $L\go\infty$ limit when $\chi_\pm$
fields aquire masses.

There are fluctuations in $Q_a$.  Write  $Q_{1,2}$=$\onehalf N \pm Q$  
and $\tQ_{1,2}$=$\onehalf N \pm \tQ$.   Important terms in $\int dx M_a
\tM_b^\dagger$ result when 
$Q=\pm \tQ$.   Since $|Q|,|\tQ| \ll N$, we have in the large volume limit
\beqn
H_{3b} = {J'\ell\over 4} \Big( {e^\gamma c \over 4\pi \hbar} \Big)^2
\int dx \hskip 9cm &&\cr
\noalign{\kern 10pt}
\times \bigg[ ~
\mu_{\Phi_-} \mu_{\chi_-} \Big\{
   e^{-i(q_1 - \tq_1)}   N[e^{-i\sqrt{4\pi}(\Phi_- + \chi_-)}]
  + e^{-i(q_2 - \tq_2)} N[e^{-i\sqrt{4\pi}(\Phi_- - \chi_-)}]   \Big\} 
\hskip 1.5cm &&\cr
\noalign{\kern 10pt}
- \mu_{\Phi_-} \mu_{\chi_+} \Big\{
   e^{-i(q_1 - \tq_2)}   N[e^{-i\sqrt{4\pi}(\Phi_- + \chi_+)}]
  + e^{-i(q_2 - \tq_1)} N[e^{-i\sqrt{4\pi}(\Phi_- - \chi_+)}]   \Big\} 
  + {\rm h.c.}  \bigg] ~. &&
\label{spinlasdder5}
\eeqn
Here we have defined $\Phi_\pm = (\Phi \pm \tPhi)/\sqrt{2}$ and
$\chi_\pm = (\chi \pm \tchi)/\sqrt{2}$.
$N[e^{-i\sqrt{4\pi}(\Phi_- + \chi_-)}]$ denotes that the $\Phi_-$ and 
$\chi_-$ fields are normal-ordered with respect to
their masses $\mu_{\Phi_-}$ and $\mu_{\chi_-}$, respectively.

$H_{3b}$ has two major effects.  It gives an additional potential in the 
zero mode sector:
\beqn
&&\Delta H_{\rm zero} = 
L{J'\ell\over 4} \Big( {e^\gamma c \over 4\pi \hbar} \Big)^2 
\mu_{\Phi_-} \cr
\noalign{\kern 10pt}
&&\hskip .5cm \times
\Big\{ \mu_{\chi_-} \big[ e^{-i(q_1 - \tq_1)}  + e^{-i(q_2 - \tq_2)} \big]
- \mu_{\chi_+} \big[ e^{-i(q_1 - \tq_2)}  + e^{-i(q_2 - \tq_1)} \big]
 + ({\rm h.c.}) \Big\} ~.
\label{spinladder6}
\eeqn
Secondly it gives additional masses to $\Phi_-$ and $\chi_\pm$.
For small $|J'| \ll J$
\beqn
\mu_{\Phi_-}^2 &=& \mu^2 - {e^{2\gamma}\over 4\pi} {J'\ell\over \hbar c}
 \, \mu_{\Phi_-} \Big( \mu_{\chi_-} \la e^{\pm i(q_1 - \tq_1)}\ra
  - \mu_{\chi_+} \la e^{\pm i(q_1 - \tq_2)} \ra \Big) \cr
\noalign{\kern 10pt}
\mu_{\chi_-}^2 &=& - {e^{2\gamma}\over 4\pi} {J'\ell\over \hbar c} \,
 \mu_{\Phi_-}  \mu_{\chi_-} \la e^{\pm i(q_1 - \tq_1)}\ra \cr
\noalign{\kern 10pt}
\mu_{\chi_+}^2 &=& ~~ {e^{2\gamma}\over 4\pi} {J'\ell\over \hbar c} \,
 \mu_{\Phi_-}  \mu_{\chi_+} \la e^{\pm i(q_1 - \tq_2)} \ra 
\label{masses1}
\eeqn
Here we have made use of 
$\la e^{\pm i(q_1 - \tq_1)}\ra = \la e^{\pm i(q_2 - \tq_2)}\ra$
and $\la e^{\pm i(q_1 - \tq_2)}\ra = \la e^{\pm i(q_2 - \tq_1)}\ra$,
which reflects the up-down symmetry of the original spin 
system and is justified shortly.

The wave function of the ladder system is specified with 
$f(p_W, \vphi; \tilde p_W, \tilde \vphi)$ as in (\ref{wavefunction1}).  
The rung interaction (\ref{spinladder6}) gives an additional potential
in the $\vphi$ representation.  $e^{iq_1}$ and $e^{iq_2}$ give rise to
$e^{i\vphi - i\pi p_W}$ and $e^{-i\vphi - i\pi p_W}$, respectively.
$f$  satisfies
\beqn
\Big\{ K(p_W, \vphi) + K(\tilde p_W, \tilde\vphi) + V_{\rm rung} \Big\}
 \, f = \ep\, f \hskip 4cm &&\cr
\noalign{\kern 10pt}
V_{\rm rung} = L^2 {J'\ell\over\pi \hbar c} 
\Big( {e^\gamma c \over 4\pi \hbar} \Big)^2  \mu_{\Phi_-}
\Big\{ \mu_{\chi_-} \cos(\vphi - \tilde\vphi) 
- \mu_{\chi_+} \cos(\vphi + \tilde\vphi) \Big\} \, 
       \cos \pi(p_W-\tilde p_W) ~. &&
\label{waveequation2}
\eeqn
For large $L$ the potential term dominates in Eq.\ (\ref{waveequation2}).
The ground state wave function has a sharp peak at the minimum of the 
potential.  For $J' >0$ ($J'<0$), the minimum occurs at
$p_W=\tilde p_W=0$ and $\vphi=-\tilde\vphi=\pm\onehalf\pi$ 
($\vphi=\tilde\vphi=\pm\onehalf\pi$) so that
\beeq
\la e^{\pm i(q_a-\tq_a)} \ra = - \la e^{\pm i(q_1-\tq_2)} \ra 
= - \la e^{\pm i(q_2-\tq_1)} \ra = \mp1   \quad
{\rm for ~} \cases{J'>0\cr J'<0~.\cr} 
\label{vev}
\eneq

The masses are determined by (\ref{masses1}) and (\ref{vev}):
\beqn
\mu_{\Phi_-} = {\mu\over \sqrt{1 - 2\kappa^2}} ~~~,~~~
\mu_{\chi_-} = \mu_{\chi_+} = {\kappa\mu\over \sqrt{1 - 2\kappa^2}} &&\cr
\noalign{\kern 10pt}
\kappa = {e^{2\gamma}\over 4\pi} \, {|J'|\ell\over \hbar c} =
{e^{2\gamma}\over 8} \, {|J'|\over J} \sim 0.397 \, {|J'|\over J}~. 
           \hskip 1.2cm &&
\label{masses2}
\eeqn
The expression is valid for small $\kappa$.  The excitation energy,
a spin gap,  is
\beeq
\Delta_{\rm spin} = \mu_{\chi_\pm} c^2 \sim \kappa\mu c^2 =
{ e^{2\gamma} k\over 4 \pi} \, |J'|
= 0.25 \, k \, |J'|~.  
\label{spingap}
\eneq
The ratio of $\Delta_{\rm spin}$ to $\mu c^2$ is $\kappa$.
The gapless mode becomes gapful. The spin gap is determined by $|J'|$,
generated irrespective of the sign of $J'$.  The energy density is lowered:
\beeq
\Delta {\cal E} = - {\Delta_{\rm spin}^2\over 2\ell J}~.
\label{energydensity}
\eneq

We have shown that the rung interaction breaks the chiral symmetry of 
spin chain systems, and generates a spin gap.  

In the literature the spin gap has been determined by various numerical
methods for varying $J'/J$.\cite{numerical}  In particular, Greven et al.\
obtained $\Delta_{\rm spin}=.41 J'$ for small $J'/J$ and $.50 J'$
for $J'=J$, which is consistent with our prediction (\ref{spingap}).

It has been well known that spin chain systems are mapped to
non-linear sigma models.\cite{Haldane}  Sierra has applied this 
mapping to $N_\ell$-leg ladder systems of spin $S$, and has shown that 
the spectrum is gapful or gapless for an integer or  half-odd-integer
$SN_\ell$, respectively.\cite{Sierra}  The mapping to sigma models is 
valid for large $SN_\ell \gg 1$, while  our method of mapping to
the Schwinger model works for $S=\onehalf$.

The method of bosonization has been employed in the spin ladder
problem.  Schulz, in analysing  a spin $S$ chain, 
expressed $\vec S$ as a sum of $2S$ spin $\onehalf$ vectors, thereby
transforming the spin chain to a special kind of a spin $\onehalf$ ladder
system. With the aid of bosonization and renormalization group analysis
he concluded that the spectrum is gapless for a half-odd-integer
$S$.\cite{Schulz}

More recently   a 2-leg $s=\onehalf$ ladder system has been analysed
by bosonization 
by Shelton et al. and by Kishine and Fukuyama.\cite{bosonize}
They have obtained a similar Hamiltonian to ours, but could not determined
the gap.  Our bosonization formula (\ref{bosonization}) is a rigorous
operator identity with no ambiguity in normalization, with which the
Hamiltonian is transformed in the bosonized form.    The correct
treatment of the normal ordering is crucial in dealing with the mass
(gap) generation.  Not only the light modes ($\chi_\pm$) but also
the heavy modes ($\Phi_\pm$) and zero modes ($\Theta, q_a$) play an
important role, which has been dismissed in ref.\ 11.

Our argument can be generalized to $N_\ell$-leg $s=\onehalf$ ladder systems. 
Inter-chain interactions are given by $H_{\rm rung}=\sum_{(ij)} J'_{ij} 
\sum_n \vec S_n^{(i)} \vec S_n^{(j)}$ where $i$ and $j$ are chain
indices and $(ij)$ labels   rung pairs.  
$J'_{ij} = 2J$ for all $(ij)$ in Schulz' model in ref.\ 10.  

Let us consider a cyclically symmetric antiferromagnetic ladder system in
which non-vanishing $J'_{ij}$'s are $J'_{i,i+1}=J'>0$ where
$J'_{N_\ell,N_\ell+1}
\equiv J'_{N_\ell,1}$.   Among boson fields
$\Phi_i$'s or $\chi_i$'s, the singlet combination is denoted by
$\Phi_+$ or $\chi_+$. Other  combinations of $\Phi$'s or $\chi$'s
are degenerate.  There are four masses to be determined:
$\mu_{\Phi_\pm}$ and  $\mu_{\chi_\pm}$.  $\mu_{\Phi_\pm} \sim \mu$
for small $|J'|$.  The issue is whether or not all
$\chi$ fields become massive.  The crucial part is the mass of $\chi_+$.

Repeating the above argument, one
finds that the part of the rung potential $V_{\rm rung}$ in
(\ref{waveequation2}),  
$\mu_{\chi_-} \cos(\vphi - \tilde\vphi) 
- \mu_{\chi_+} \cos(\vphi + \tilde\vphi) $,
is replaced by
\beeq
\mu_{\chi_-} \sum_{i=1}^{N_\ell} \cos(\vphi_i - \vphi_{i+1}) 
- \mu_{\chi_+}^{2/N_\ell} \mu_{\chi_-}^{1-(2/N_\ell)} 
 \sum_{i=1}^{N_\ell} \cos(\vphi_i + \vphi_{i+1})  
\label{potential2}
\eneq
where $\vphi_{N_\ell+1}=\vphi_1$.  
If $\mu_{\chi_-}=0$, $V_{\rm rung}=0$ 
and no correction arises to $\mu_{\Phi_\pm}$ or  $\mu_{\chi_\pm}$.
This solution has a higher energy density than
the non-trivial solution so that $\mu_{\chi_-}\not=0$.  
From the symmetry  $V_{\rm rung}$
is minimized at $\cos(\vphi_i - \vphi_{i+1}) = f_-$ ($i=1,\cdots,N_\ell$).
This implies that $\vphi_j=\vphi + (j-1)\eta$ and $\eta=2p\pi/N_\ell$ or
$2p\pi/(N_\ell-2)$ where $p$ is an integer. 

Suppose  $\mu_{\chi_+}\not= 0$.
Then $\cos(\vphi_i + \vphi_{i+1}) = f_+$ ($i=1,\cdots,N_\ell$).  This leads
to an additional  condition that $\eta=\pi$.
All of these conditions are satisfied for an even $N_\ell$.  
The potential is minimized at $\vphi_{2p+1}=\pm \onehalf \pi$ and
$\vphi_{2p}=\mp \onehalf\pi$.   For an odd $N_\ell$  the conditions cannot
be satisfied.

If $\mu_{\chi_+} = 0$, $\eta$ need not be $\pi$.  This gives a solution
for an odd $N_\ell$.  For an even $N_\ell$, this solution yields a higher energy
density than the solution with $\mu_{\chi_+} \not= 0$ above.
To summarize,  the spectrum is gapless for an odd $N_\ell$,
but is gapful for an even $N_\ell$.   The interaction is frustrated in
the  rung direction for an odd $N_\ell$.  The argument here is similar to 
Schulz' in ref.\ 10.

In the experimental samples\cite{exp3} $J' \sim J$ so that
$\kappa={\rm O}(1)$. For instance, in SrCu$_2$O$_3$ (2-leg ladder), 
$J\sim J'\sim 1300$K and  $\Delta_{\rm spin}\sim 420$K.
The formula (\ref{masses1}) need to be improved by taking account of 
effects of nonlinear terms in (\ref{spinlasdder5}).  Further it is observed
that spin ladder systems with three legs are gapless.  
[The experimental sample is not cyclically symmetric:  $J'_{12}=J'_{23}
\sim J$ but  $J'_{13}=0$.] 
For this the large value of $\kappa$ is important, as our analysis
indicates that a gap is generated so long as $\kappa$ is sufficiently small.   
It has been also reported that
the spin gap is not affected by nonmagnetic impurities.\cite{exp4}  We
will come back to those points in separate publications.

\vskip .5cm 

\leftline{\bf Acknowledgment}
This work was supported in part  by the U.S.\ Department of Energy
under contracts DE-FG02-94ER-40823.

\def\jnl#1#2#3#4{{#1}{\bf #2} (#4) #3}

\def\em{\it}
\def\nc{\em Nuovo Cimento }
\def\jpA{{\em J.\ Phys.} A}
\def\jpC{{\em J.\ Phys.\ Cond.\ Mat.} }
\def\npB{{\em Nucl.\ Phys.} B}
\def\plA{{\em Phys.\ Lett.} A}
\def\plB{{\em Phys.\ Lett.} B}
\def\prl{\em Phys.\ Rev.\ Lett. }
\def\pr{{\em Phys.\ Rev.} }
\def\prB{{\em Phys.\ Rev.} B}
\def\prD{{\em Phys.\ Rev.} D}
\def\ap{{\em Ann.\ Phys.\ (N.Y.)} }
\def\rmp{{\em Rev.\ Mod.\ Phys.} }
\def\zpC{{\em Z.\ Phys.} C}
\def\sci{\em Science}
\def\cmp{\em Comm.\ Math.\ Phys. }
\def\mplA{{\em Mod.\ Phys.\ Lett.} A}
\def\mplB{{\em Mod.\ Phys.\ Lett.} B}
\def\ijmpB{{\em Int.\ J.\ Mod.\ Phys.} B}
\def\IJMPB{{\em Int.\ J.\ Mod.\ Phys.} B}
\def\ijmpA{{\em Int.\ J.\ Mod.\ Phys.} A}
\def\IJMPA{{\em Int.\ J.\ Mod.\ Phys.} A}
\def\ptp{{\em Prog.\ Theoret.\ Phys.} } 
\def\Zphys{{\em Z.\ Phys.} }
\def\jpsJ{{\em J.\ Phys.\ Soc.\ Japan }}
\def\jmp{{\em J.\ Mod.\ Phys.} }
\def\jssc{{\em J.\ Solid State Chem.\ }}

\def\etal{{\em et al,} }

\vskip .5cm
\leftline{\bf References}

\renewenvironment{thebibliography}[1]
	{\begin{list}{[\arabic{enumi}]}
	{\usecounter{enumi}\setlength{\parsep}{0pt}
	 \setlength{\itemsep}{0pt}
         \settowidth
 {\labelwidth}{#1.}\sloppy}}{\end{list}}

\myend